\documentclass[11pt,letterpaper]{article}
\usepackage{lineno}
\usepackage[figuresonly,nomarkers,nolists]{endfloat}
\usepackage[T1]{fontenc}
\usepackage[utf8]{inputenc}
\usepackage{lmodern}
\usepackage[margin=1in]{geometry}
\usepackage{microtype}
\usepackage{graphicx}
\usepackage{amsmath,amssymb}
\usepackage{textcomp}
\usepackage{xcolor}
\usepackage{authblk}
\usepackage{caption}
\usepackage{csquotes}
\usepackage[backend=biber,style=chem-acs,sorting=none,doi=true,url=false,isbn=false]{biblatex}
\usepackage[hidelinks]{hyperref}

\title{Morphology-Guided Deterministic Fabrication of Low-Noise High-Temperature Superconducting Quantum Interference Devices}

\date{}
\author[1,$\dagger$, *]{Bingke Xiang}
\author[2,$\dagger$]{Wanjuan Tang}
\author[2]{Shiqun Liu}
\author[1]{Lingtong Hou}
\author[2]{Geming Zhang}
\author[1]{Yibo~Wang}
\author[1]{Ruonan~Wang}
\author[1]{Zhiqiang~Cao}
\author[2]{Jiaqi Wei}
\author[3,*]{Xueshen~Wang}
\author[1,2,4]{Xueying~Zhang}
\author[1,2,4,*]{Xiaoyang Lin}

\affil[1]{State Key Laboratory of Spintronics, Hangzhou International Innovation Institute, Beihang University, Hangzhou 311115, China}
\affil[2]{Fert Beijing Institute, School of Integrated Circuit Science and Engineering, Beihang University, Beijing 100191, China}
\affil[3]{National Institute of Metrology, Beijing 100029, China}
\affil[4]{Truth Instruments Co. Ltd., Qingdao 266100, China}
\affil[*]{Address correspondence to:
\mbox{\href{mailto:xiangbk@buaa.edu.cn}{\textnormal{xiangbk@buaa.edu.cn}}},
\mbox{\href{mailto:wangxs@nim.ac.cn}{\textnormal{wangxs@nim.ac.cn}}}, and
\mbox{\href{mailto:XYLin@buaa.edu.cn}{\textnormal{XYLin@buaa.edu.cn}}}}
\affil[$\dagger$]{These authors contributed equally to this work.}

\begin{document}
\maketitle

\begin{center}
\end{center}
\medskip
\begin{abstract}
\fontsize{10}{12}\selectfont
Reproducible bicrystal high-temperature superconducting quantum interference
devices remain limited by local variability along the grain
boundaries that form the Josephson junctions. Here, we develop a site-selective fabrication workflow in which atomic force microscopy maps
the intended junction region before lithography, quantifies an apparent
grain-boundary width, rejects pore-rich segments, and writes a nearby
registration mark for site-specific pattern alignment. The apparent grain-boundary width provides a practical morphology metric, with narrower regions consistently yielding larger critical currents and characteristic voltages. Iterative optimization within this workflow further improves junction and device performance, reaching a liquid-nitrogen-temperature field-noise level of 40 fT Hz$^{-1/2}$. This strategy turns local grain-boundary heterogeneity from an uncontrolled source of variability into a basis for site-selective fabrication, providing a route towards scalable manufacturing of low-noise HTS SQUIDs with high uniformity.

\end{abstract}

\newpage
\par\medskip
High-temperature superconducting quantum interference devices offer
ultrasensitive magnetic detection near the boiling point of liquid
nitrogen, reducing cryogenic complexity relative to
low-temperature-superconductor systems, enabling compact sensors for
biomagnetism, nondestructive evaluation, geophysics, and magnetic
microscopy.\supercite{koelle1999,hilgenkamp2002,Kober2016SUSTreviewBio,Clarke2018SUSTreviewBio,Stolz2022MineralReview,Kirtley1999Review,Xiang2024EMreview} Their performance is determined
by two Josephson junctions whose critical currents, resistances,
asymmetry, and low-frequency fluctuations set the transfer function and
noise. Bicrystal grain boundaries in high-temperature superconductors are among the most
established junction technologies because the substrate fixes a nominal
crystallographic misorientation and the film reproduces the boundary
during epitaxial growth.\supercite{koelle1999,hilgenkamp2002,tafuri2005}

The nominal misorientation angle controls the mean suppression of the
grain-boundary critical current, as established by early bicrystal
studies.\supercite{dimos1988,mannhart1988,graser2010} It does not uniquely determine the local
junction formed at a particular lithographic coordinate. Grain-boundary
dislocation structure, faceting, strain, oxygen stoichiometry, secondary
phases, thickness modulation, and the three-dimensional trajectory of
the film boundary can all vary along a single macroscopic
interface.\supercite{hilgenkamp2002,chisholm1991,ayache1998,traeholt1994} Consequently, junctions patterned with
the same design and on the same nominal bicrystal can exhibit markedly
different electrical characteristics. Conventional photolithography
treats the grain boundary as a uniform line and therefore samples this
local heterogeneity largely by chance.

Previous studies have established that local grain-boundary morphology can strongly influence the properties of bicrystal Josephson junctions and SQUIDs. AFM and electron-microscopy studies correlated groove-like boundary defects with junction transport and SQUID performance, including reduced flux noise for junction regions with shallow grooves.\supercite{yu1999,wu2006} These results highlight the importance of local morphology, but leave a key fabrication challenge unresolved: how to identify favorable junction sites quantitatively and transfer the selected microscopic locations reliably into subsequent device patterning. Improving the structural quality of bicrystal substrates offers a complementary upstream strategy,\supercite{ding2020} but cannot eliminate local variations that emerge in the superconducting film after growth. Direct inspection and selection of the actual device-forming boundary therefore remain necessary.

Here, we establish an AFM-guided, site-selective fabrication workflow that combines local morphology assessment, quantitative width analysis, and probe-written registration. The apparent full width at half-maximum of the boundary-associated topographic depression, denoted $w_{\mathrm{AFM}}$, is used to rank candidate junction sites. A nearby AFM-written mark preserves the coordinates of selected regions for subsequent registered lithography, enabling junctions to be fabricated at preferred locations along the bicrystal boundary. Using this strategy, we correlate local morphology with Josephson transport and progressively optimize junction and SQUID performance. Importantly, $w_{\mathrm{AFM}}$ is not interpreted as a microscopic weak-link or barrier length, but rather as an operational, morphology-derived process-control metric. By turning intrinsic grain-boundary variability into a selectable fabrication degree of freedom, this approach provides a practical route towards improved yield, reproducibility and performance while retaining the established bicrystal-junction architecture. More broadly, it establishes a materials-informed fabrication paradigm in which nanoscale heterogeneity is incorporated into device placement before patterning, rather than treated only as a source of variability after fabrication.

\section*{METHODS}

 Bicrystal substrates and HTS films were examined by atomic force microscopy (AFM) prior to device patterning. Local grain-boundary morphology was evaluated from AFM topographs, with continuous boundary segments and localized pore-like defects identified for site selection. The apparent grain-boundary width, $w_{\mathrm{AFM}}$, was defined as the full width at half-maximum (FWHM) of the resulting profile. This quantity was used as an empirical morphology descriptor rather than a measure of the microscopic weak-link width.

Following AFM inspection, a local probe-written mark adjacent to the selected grain-boundary region was used as a coordinate reference for subsequent lithographic alignment and device patterning. Josephson junctions and dc SQUIDs were fabricated across the bicrystal grain boundary using standard microfabrication procedures.

Electrical characterization included current--voltage, temperature-dependent resistance, and magnetic-field-dependent critical-current measurements. Effective $I_c(B)$ values were extracted using resistively shunted junction (RSJ) fits. SQUID noise spectra were measured under both dc- and ac-bias operation at liquid-nitrogen temperature.

\section*{RESULTS AND DISCUSSION}

\subsection*{Morphology-Guided Selection and Local Registration}

Figure 1 illustrates the principle of morphology-guided site selection. The superconducting film
inherits a grain boundary whose apparent width and defect population
vary along its length (Fig. 1a). Patterning a junction at an arbitrary
coordinate therefore conflates the designed bridge geometry with an
uncontrolled local interface. In the proposed workflow, atomic force
microscopy is first used to locate a narrow, continuous segment without
resolved pore-like defects. A straight probe-written mark is then placed
adjacent to the selected segment to preserve the
electrical path while providing an unambiguous local reference (Fig. 1b). Registered lithography subsequently aligns the
superconducting-quantum-interference-device loop so that its two
junctions intersect the selected part of the boundary (Fig. 1c).

The alignment step addresses a practical metrology gap between nanoscale characterization and microfabrication. A favourable GB segment identified within the field of view of an AFM image is otherwise difficult to re-identify with the positional fidelity required during subsequent mask alignment or direct-write lithography. The AFM-written mark provides a common spatial reference between these otherwise distinct coordinate systems, preserving the location selected during morphology screening through the subsequent fabrication sequence.

The selection step can be performed on the as-grown film,
before the irreversible and comparatively expensive sequence of resist
coating, exposure, etching, dicing, packaging, and cryogenic characterization. It allows unfavourable regions to be rejected before substantial downstream processing is committed. The approach therefore changes local GB morphology from a largely uncontrolled fabrication variable into an experimentally accessible criterion for pre-fabrication site selection. Lithography still determines the designed junction geometry, but AFM screening additionally controls which local interfacial region that geometry samples. In this sense, morphology-guided site selection adds a materials-level process-control dimension to an otherwise predominantly geometry-defined fabrication process.

\begin{figure}[htbp]
\centering
\includegraphics[width=\linewidth]{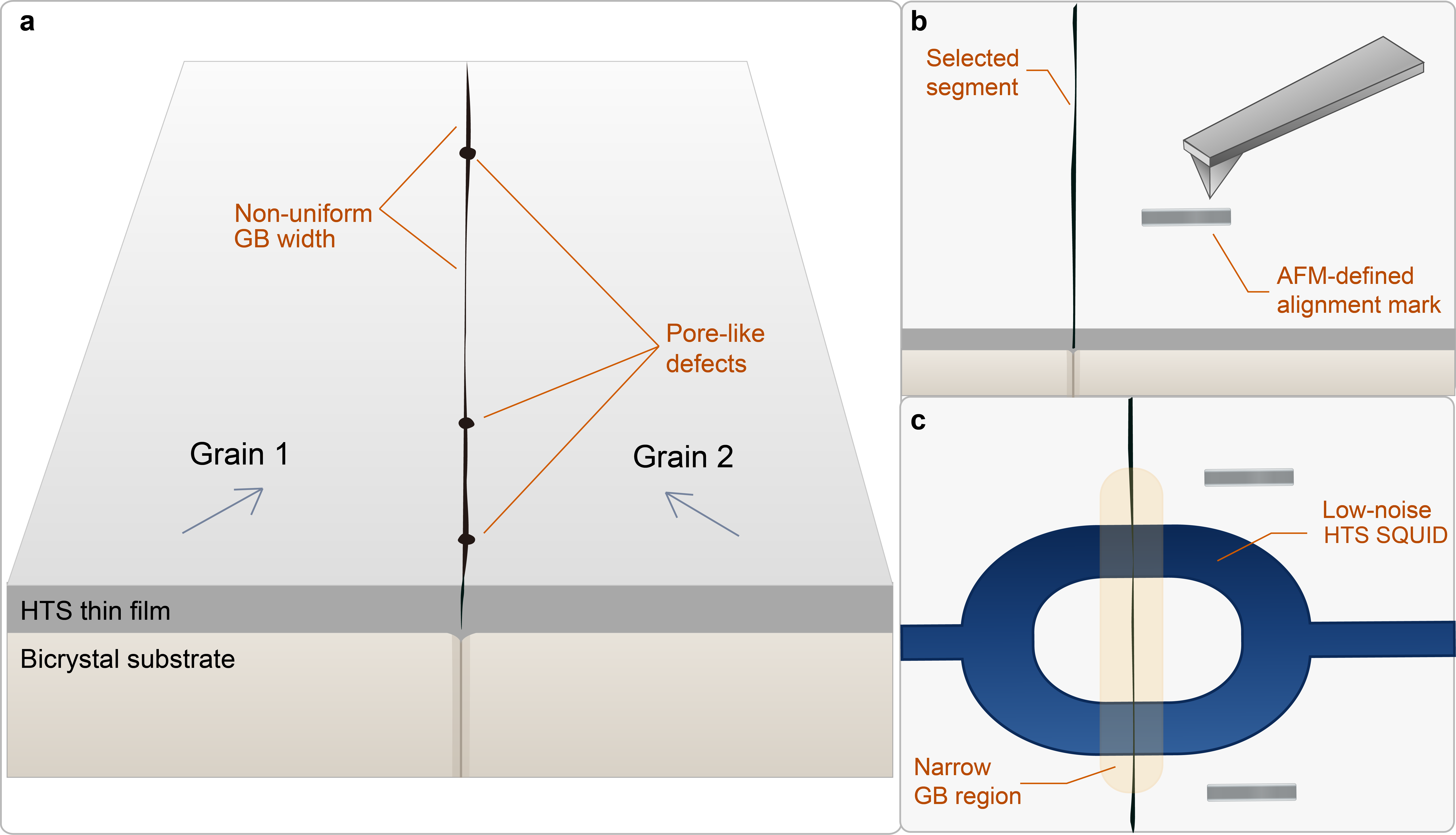}
\caption{AFM-guided selection of local grain-boundary regions for HTS SQUID fabrication. \textbf{a,} Schematic of a high-temperature superconducting (HTS) thin film on a bicrystal substrate, illustrating spatial variations in the apparent grain-boundary width and local pore-like defects. \textbf{b,} Atomic force microscopy (AFM) inspection identifies a locally narrow segment without visible pore-like defects, and an adjacent AFM-defined alignment mark provides a local reference for subsequent lithography. \textbf{c,} Registered patterning places the two Josephson junctions of the dc superconducting quantum interference device (SQUID) across the selected narrow grain-boundary region.}
\label{fig:figure1}
\end{figure}

\subsection*{Atomic-Force-Microscopy-Resolved Boundary Heterogeneity}

The topographs in Figure 2 show that nominally continuous bicrystal
boundaries contain at least two distinct classes of surface
heterogeneity. In the pore-rich region (Fig. 2a), the boundary trace
is accompanied by spatially nonuniform depressions, and the profile
crossing a representative pore is substantially broader than the profile
crossing the adjacent continuous boundary. In a region without resolved
pores (Fig. 2b), the boundary remains visible but its apparent width
changes continuously over micrometer distances. The straight mark in
Fig. 2c demonstrates that a local reference can be produced adjacent
to the chosen position without intentionally interrupting the boundary.

To compare these features, profiles normal to the boundary were aligned
to their minima, normalized by depression depth, and represented by an
apparent full width at half-maximum (Fig. 2d). The displayed examples
span a narrow continuous segment of roughly 60--70 nm, an ordinary
boundary segment near 100 nm, a laterally wider segment about 130 nm, and
a pore-like depression around 160 nm. 

\begin{figure}[htbp]
\centering
\includegraphics[width=\linewidth]{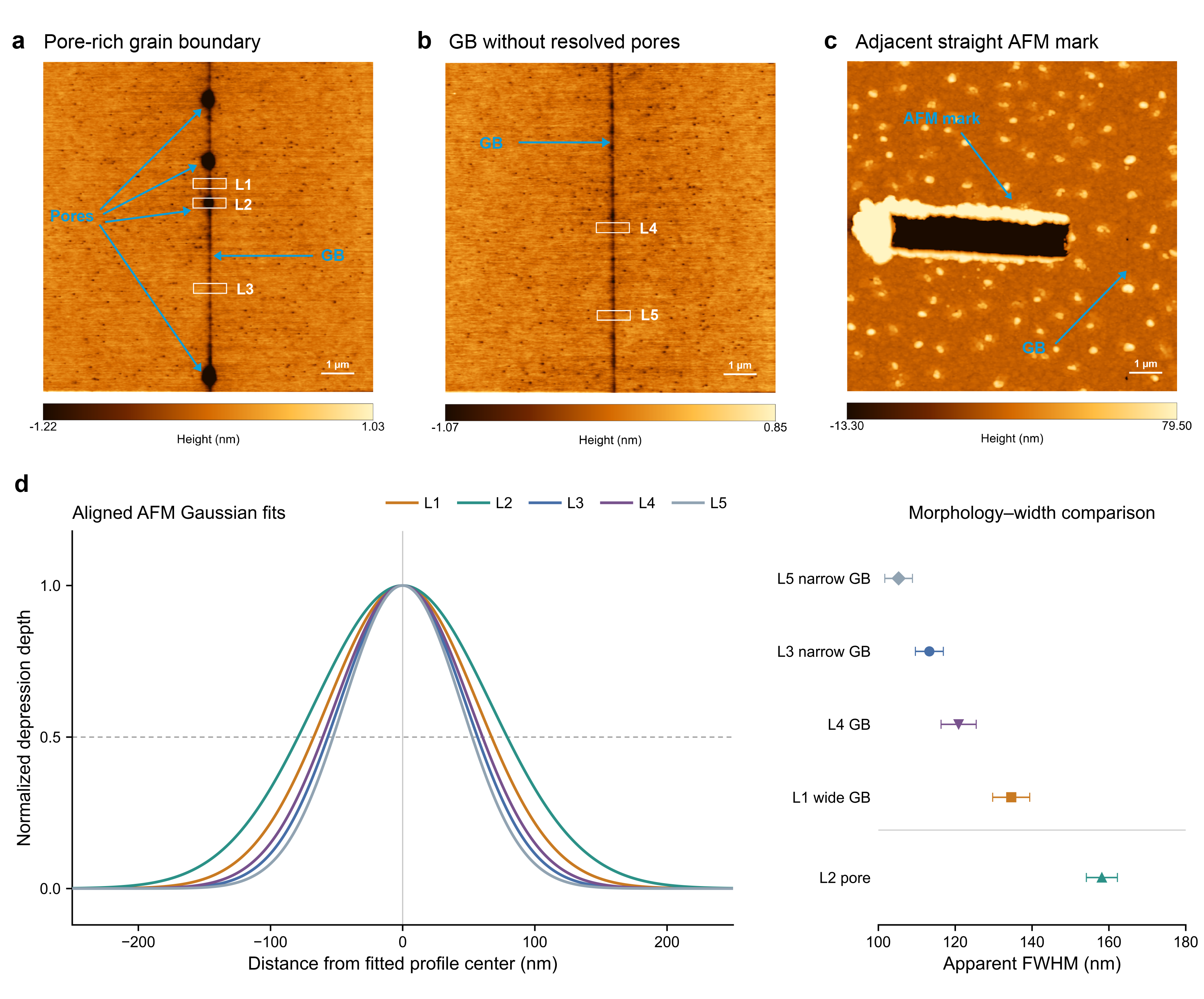}
\caption{AFM-resolved grain-boundary heterogeneity and local alignment mark. \textbf{a,} AFM topography of a pore-rich bicrystal grain boundary (GB). Line profiles L1 and L3 intersect the GB at two local positions, whereas L2 traverses a representative pore-like depression. \textbf{b,} AFM topography of a region without resolved pore-like defects; L4 and L5 sample segments with different apparent widths. \textbf{c,} AFM-written mark adjacent to the selected GB region, providing a local fiducial feature for subsequent lithographic alignment and site-specific junction patterning. \textbf{d,} Left, averaged line profiles from L1--L5, aligned to their minima and normalized by depression depth. Right, corresponding apparent full widths at half maximum (FWHM), providing the operational AFM-resolved width used in subsequent analyses. Horizontal error bars denote the standard errors of the fitted FWHM values.}
\label{fig:figure2}
\end{figure}

Atomic force microscopy records
the free-surface morphology $w_{\mathrm{AFM}}$ , which is not the Josephson-barrier thickness or a direct measure of a coherence length. Instead, it is an empirically accessible descriptor
that may integrate several underlying factors, including thickness
depression, faceting and strain
accommodation. The practical
question below is whether this composite descriptor contains
enough information to rank fabrication sites.

\subsection*{Josephson Junction Transport in Relation to Apparent Boundary Width}

\begin{figure}[htbp]
\centering
\includegraphics[width=\linewidth]{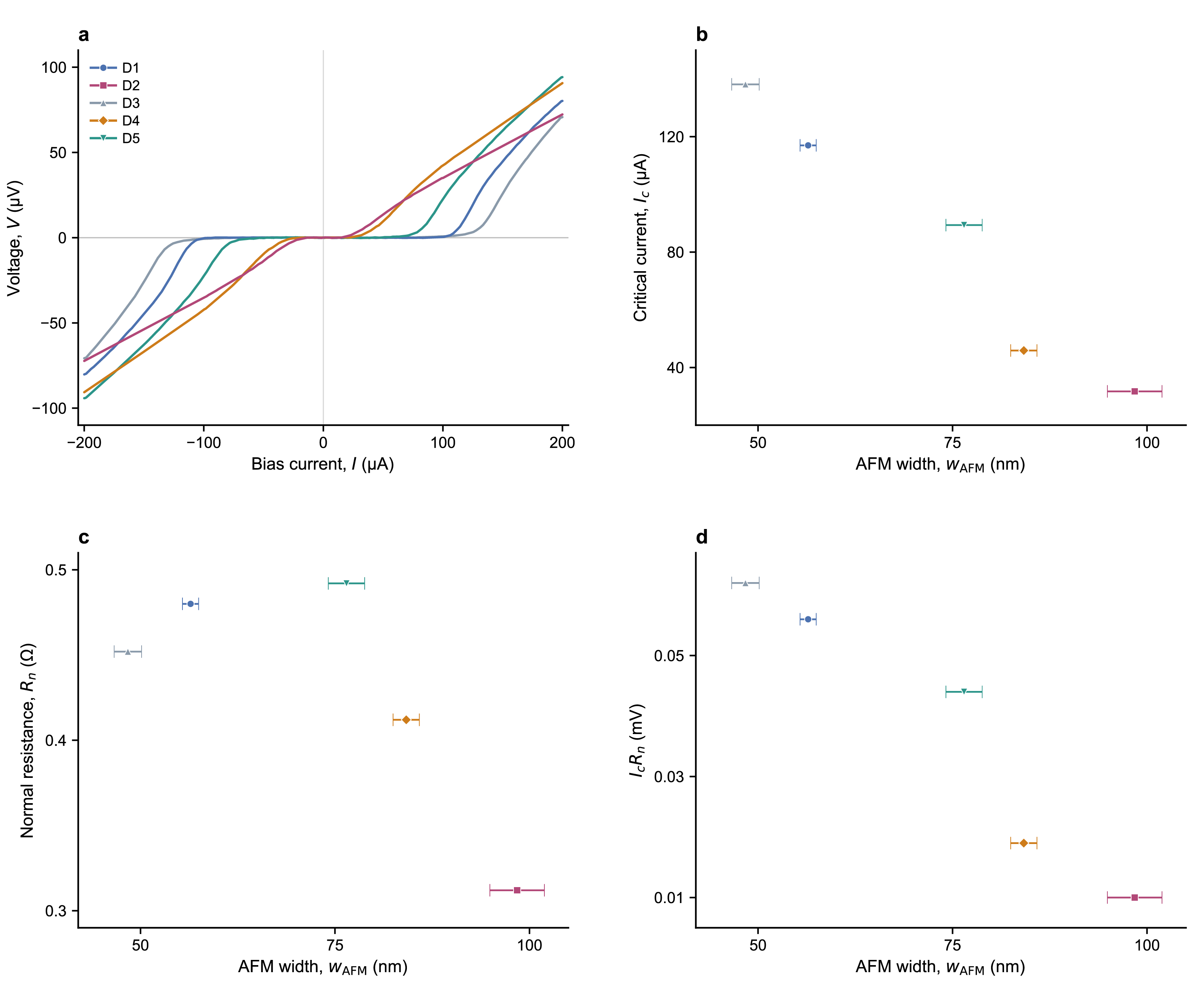}
\caption{Systematic variation of Josephson junction transport with AFM-resolved grain-boundary width. \textbf{a,} Current--voltage characteristics of bicrystal Josephson junction devices D1--D5. \textbf{b--d,} Critical current, $I_{\mathrm{c}}$ (\textbf{b}), normal-state resistance, $R_{\mathrm{n}}$ (\textbf{c}), and characteristic voltage, $I_{\mathrm{c}}R_{\mathrm{n}}$ (\textbf{d}), plotted against the AFM-resolved grain-boundary width, $w_{\mathrm{AFM}}$. $I_{\mathrm{c}}$ decreases markedly with increasing $w_{\mathrm{AFM}}$, whereas $R_{\mathrm{n}}$ exhibits a comparatively weaker width dependence. Colors identify the same devices across all panels, with corresponding symbol shapes used in b--d. Horizontal error bars indicate the standard errors of the fitted AFM-derived width.}
\label{fig:figure3}
\end{figure}

Five bicrystal Josephson junctions fabricated at positions with
different $w_{\mathrm{AFM}}$ values exhibit current--voltage
characteristics with a clear spread in critical current (Fig. 3a).
Across the set, $w_{\mathrm{AFM}}$ increases from approximately 40 to 100 nm, whereas
$I_{\mathrm{c}}$ decreases from approximately 140 to 31 $\mu$A (Fig. 3b). The narrowest position supports the largest values $I_{\mathrm{c}}$, whereas the position near 100 nm exhibit substantially smaller critical currents. The intermediate-width device follows the same overall ordering. These observations establish a clear association between the local AFM-resolved grain-boundary morphology and junction critical current within the present device set.


In contrast, the normal-state resistance varies over a comparatively narrow range of approximately 0.31--0.49 $\Omega$, and shows no monotonic dependence on $w_{\mathrm{AFM}}$ (Fig. 3c).
Consequently, the characteristic voltage $I_{\mathrm{c}}R_{\mathrm{n}}$ decreases strongly, from
approximately 0.063 mV at the narrow end of the data set to
approximately 0.010 mV for the widest and weakest junction (Fig. 3d).
The dominant trend therefore enters through $I_{\mathrm{c}}$ rather than through a
compensating increase in $R_{\mathrm{n}}$. 


Taken together, these results identify $w_{\mathrm{AFM}}$ as a useful pre-fabrication morphology metric for ranking candidate junction sites. Importantly, this correlation does not imply that $w_{\mathrm{AFM}}$ represents the microscopic weak-link or barrier length; rather, it provides an operational measure of the local surface morphology associated with the grain boundary. The present five-device data set is sufficient to reveal a clear ordering of junction performance with $w_{\mathrm{AFM}}$, but not to establish a universal functional dependence or a sharp physical threshold. Measurements across larger device populations, wafers and fabrication batches will be required to quantify the statistical robustness and process-to-process transferability of this relationship.

For the present fabrication process, the data nevertheless support a provisional morphology-based ranking strategy. Regions containing resolved pore-like defects are excluded during AFM screening. Among continuous grain-boundary segments, sites with $w_{\mathrm{AFM}}$ below 60 nm are preferentially selected, regions in the approximately 60–80 nm range are treated as intermediate candidates, and substantially wider regions, approaching 90–100 nm, are assigned lower priority. These ranges should be regarded as process-specific screening bands rather than universal material thresholds. Their practical value lies in enabling candidate sites to be ranked before lithography and cryogenic characterization, with the selection criteria refined as additional device statistics become available.

\subsection*{From Junction Heterogeneity to SQUID Noise}

Independent measurements at the junction and circuit levels provide complementary signatures of device-to-device heterogeneity (Figure 4). The normalized $R(T)$ characteristics of junctions J1–J3 exhibit distinct transition widths and resistive tails (Fig. 4a), indicating that superconducting transport across the junction region is established over different temperature intervals in different devices. Resistive tails in YBCO structures have previously been associated with weakened grain-boundary coupling, while the transport properties of bicrystal boundaries are known to be highly sensitive to local structural disorder. The observed variation is therefore consistent with differences in the effective weak-link environment sampled by individual junctions, although $R(T)$ alone cannot uniquely identify the microscopic origin of this inhomogeneity.

The magnetic interference response provides a complementary probe of the spatial uniformity of Josephson coupling (Fig. 4b). For an ideal short junction with a uniform current density, a regular Fraunhofer-like pattern is expected; deviations therefore encode a
nonuniform effective current distribution.\supercite{dynes1971} The
present patterns are consistent with spatially heterogeneous
supercurrent transmission\supercite{rosenthal1991,mayer1993,gerdemann1994,carmody2000,camerlingo2002}. They do not uniquely reconstruct microscopic
conducting channels, because flux focusing, field misalignment, junction
faceting, phase shifts, finite junction length, and the loss of phase
information in \textbar $I_{\mathrm{c}}(B)$\textbar{} can produce nonunique
inversions. We use the interference patterns primarily as a diagnostic of junction uniformity.

The consequences of junction variability become particularly evident in the SQUID noise spectra (Figs. 4c,d). Under direct-current bias,
low-frequency excess noise extends across much of the measured band.
Alternating-current operation flattens the spectrum 
and suppresses the low-frequency contribution, consistent with
established strategies for reducing critical-current-fluctuation noise
in high-temperature superconducting devices.\supercite{dantsker1996,oukhanski2003} The
best device reaches a field-noise level of 40 fT Hz$^{-1/2}$
at liquid-nitrogen temperature. Together, these measurements connect junction-level variability with the broader spread in circuit-level SQUID performance, while a direct device-by-device correlation with $w_{\mathrm{AFM}}$ remains to be established.



\begin{figure}[htbp]
\centering
\includegraphics[width=\linewidth]{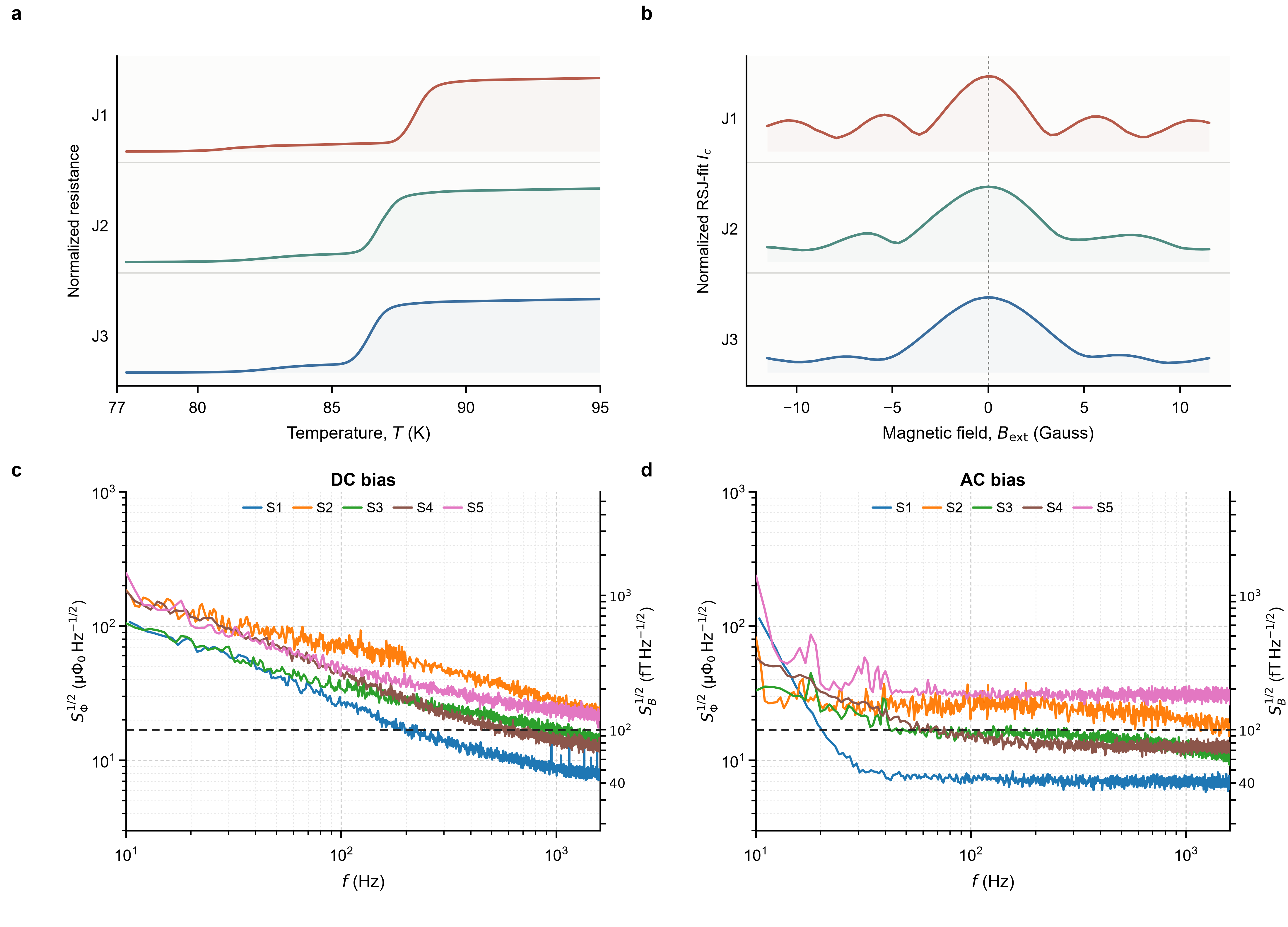}
\caption{Signatures of local grain-boundary heterogeneity in Josephson transport and SQUID noise. \textbf{a,} Normalized resistance as a function of temperature for three bicrystal Josephson junctions (J1--J3), vertically offset for clarity. The junctions exhibit distinct transition broadening and resistive tails. \textbf{b,} Normalized effective critical current $I_{\mathrm{c}}(B)$ extracted from resistively shunted junction (RSJ) fits. Traces are vertically offset. The device-specific lobe structures are consistent with spatially non-uniform effective supercurrent transmission across the grain boundary. \textbf{c,d,} Noise amplitude spectral densities for five bicrystal SQUIDs (S1--S5) under DC and AC bias, respectively. Right axes show the equivalent field noise; the dashed line marks $100\,\mathrm{fT}\,\mathrm{Hz}^{-1/2}$.}
\label{fig:figure4}
\end{figure}

The noise achieved is competitive with established low-noise HTS SQUIDs, for
which interface quality, resistive shunting, geometry, and readout
strategy are decisive \supercite{Beyer1998SUST,nagel2011,schwarz2013,Xu2023pkubicrystal100fT,lin2020,trabaldo2019,Ruffieux2020SUST}. The distinguishing
manufacturing contribution here is upstream site selection: rather than
relying only on a nominal bicrystal angle and postfabrication electrical
sorting, the workflow achieves to select favorable local boundary
regions before junction definition.

\subsection*{Process-Control Implications and Scalability}

The proposed approach provides coordinate-level process control by using reproducible surface morphology to rank candidate junction sites before device patterning. This preselection increases the likelihood that both junctions are formed within a favorable parameter range, thereby reducing junction asymmetry and improving device consistency and yield. Because screening is performed upstream of lithography and cryogenic testing, unfavorable regions can be excluded before costly downstream processing.

Importantly, this approach does not require exhaustive AFM mapping of the entire grain boundary. Instead, AFM screening can be restricted to a limited number of candidate junction regions identified before device patterning, substantially reducing the area and time required for high-resolution characterization. The workflow is also amenable to further automation: grain-boundary localization, targeted AFM imaging, profile analysis and site ranking could be integrated with local alignment-mark writing and coordinate transfer to subsequent lithography. Such a hierarchical screening-and-registration strategy provides a practical route towards higher-throughput and more reproducible fabrication while retaining the site-specific control required for high-performance HTS SQUIDs.


\section*{CONCLUSIONS}

Local grain-boundary morphology emerges from this work as a practical process variable for controlling bicrystal Josephson-junction fabrication. AFM reveals substantial local heterogeneity along nominally identical boundaries, while the apparent boundary width provides a quantitative means to rank candidate sites: narrower, continuous regions are associated with superior junction transport, whereas pore-like defects identify unfavorable locations. Combined with local probe-written registration, this information enables preferred grain-boundary regions to be selected and transferred directly into subsequent device fabrication.

This approach shifts bicrystal SQUID fabrication from random junction placement and post-fabrication sorting toward measurement-guided, site-selective manufacturing. Progressive refinement of junction quality within this workflow ultimately yields high-quality Josephson junctions and low-noise HTS SQUIDs. More significantly, converting intrinsic grain-boundary heterogeneity from an uncontrolled source of device variability into an actionable fabrication parameter provides a foundation for scalable and reproducible manufacturing of high-performance HTS SQUIDs with improved device-to-device uniformity.

\section*{ACKNOWLEDGMENTS}

This work was supported by the National Key R\&D Program of China (2023YFF0720500) and the Research Startup Funds of Hangzhou International Innovation Institute of Beihang University (2025BKZ030).

\section*{DATA AVAILABILITY}

The data and analysis code supporting this study are available from the corresponding author upon reasonable request.

\printbibliography[title={REFERENCES}]

\clearpage

 \end{document}